\documentclass[iicol,pdflatex,sn-apa]{sn-jnl}

\usepackage{graphicx}%
\DeclareGraphicsExtensions{.eps}
\usepackage{multirow}%
\usepackage{amsmath,amssymb,amsfonts}%
\usepackage{mathtools}%
\usepackage{amsthm}%
\usepackage{mathrsfs}%
\usepackage[title]{appendix}%
\usepackage{xcolor}%
\usepackage{textcomp}%
\usepackage{manyfoot}%
\usepackage{booktabs}%
\usepackage{algorithm}%
\usepackage{algpseudocode}%
\usepackage{listings}%

\usepackage{latexsym}
\usepackage{float}
\usepackage{subfig}
\usepackage[]{placeins}
\usepackage{multirow}
\usepackage[subnum]{cases}
\usepackage{tabularx}

\usepackage{hyperref}
\usepackage{subfig}

\theoremstyle{thmstyleone}%
\theoremstyle{thmstyletwo}%

\theoremstyle{thmstylethree}%

\begin{document}

\title[IDSS, a Novel P2P Relational Data Storage Service]{IDSS, a Novel P2P Relational Data Storage Service}


\author*[1]{\fnm{Massimo} \sur{Cafaro}}\email{massimo.cafaro@unisalento.it}

\author[1]{\fnm{Italo} \sur{Epicoco}}\email{italo.epicoco@unisalento.it}

\author[1]{\fnm{Marco} \sur{Pulimeno}}\email{marco.pulimeno@unisalento.it}

\author[1]{\fnm{Lunodzo J.} \sur{Mwinuka}}\email{lunodzo.mwinuka@unisalento.it}

\author[2]{\fnm{Lucas} \sur{Pereira}}\email{lucas.pereira@tecnico.ulisboa.pt}

\author[2]{\fnm{Hugo} \sur{Morais}}\email{hugo.morais@tecnico.ulisboa.pt}

\affil*[1]{\orgdiv{Department of Engineering for Innovation}, \orgname{University of Salento}, \orgaddress{\city{Lecce}, \postcode{73100}, \country{Italy}}}

\affil[2]{\orgdiv{Instituto Superior Técnico}, \orgname{Universidade de Lisboa}, \orgaddress{\city{Lisbon}, \postcode{1049-001}, \country{Portugal}}}
%


\abstract{The rate at which data is generated has been increasing rapidly, raising challenges related to its management. Traditional database management systems suffer from scalability and are usually inefficient when dealing with large-scale and heterogeneous data. This paper introduces IDSS (InnoCyPES Data Storage Service), a novel large-scale data storage tool that leverages peer-to-peer networks and embedded relational databases. We present the IDSS architecture and its design, and provide details related to the implementation. The peer-to-peer framework is used to provide support for distributed queries leveraging a relational database architecture based on a common schema. Furthermore, methods to support complex distributed query processing, enabling robust and efficient management of vast amounts of data are presented.}




\maketitle

\section{Introduction}
\label{sec:intro}
Data drives decision-making, and, owing to its importance, it is currently the most valuable asset in the world \cite{RN81}. Little wonder most of the world's prestigious companies are gunning for it and harnessing the power of data to improve their products, take strategic decisions, refine operations, and create new income streams. As communities strive for efficiency, sustainability, and innovation, the need for connectivity and the number of connected devices is increasing. This further fuels an increase in data that is generated and processed. The data generated by these devices (commonly used in the industries, such as Internet of Things (IoT) devices, smart grids, smart meters, etc.) are generated at a very fast pace: indeed, some are generated every few milliseconds.

The main problem is that rapidly growing data could be in different formats, distributed across multiple platforms, and require re-purposing \cite{RN80}. Hence, there is a need to manage the data, store crucial information for reuse, and mine that data to infer new knowledge and gain insights. Data that come in different formats further make its integration and fusion for more robust comprehension very tricky, as the process needs to be done in a way that guarantees the information to be consistent, accurate, and secure \cite{RN66}.

Developments in the management, utilisation, and analysis of such large datasets have emerged to address rising challenges. This is witnessed by a shift from traditional relational database technologies to NoSQL, due to NoSQL's increased flexibility and scalability \cite{RN46,RN152,RN161,RN63}. In some cases, Machine Learning (ML) models have been employed to facilitate data fusion \cite{RN127, RN66}. Unfortunately, an automated approach to data processing cannot, at least currently, be perfect. Hence, the need for human intervention to some extent is still crucial \cite{sundsgaard2025data, RN278}, and starts in the early stages of data collection and storage before moving on to the later stages of data retrieval and processing. Also, managing data with NoSQL database systems, despite adding an advantage to flexibility in data modelling, if deployed in a centralised architecture, adds limitations in terms of scalability and limited fault tolerance, and introduces a single point of failure. Another challenge currently prevailing is data availability and access, because most digital devices are operating across distributed areas, and each one is generating its own data. In this regard, information systems are collecting more data than ever, but most of it remains idle or stuck in siloed storage, with significant untapped potential \cite{RN45}.

Decentralised data storage is an important aspect of data management, as it aids the entire data science lifecycle. It plays a vital role in data analysis and processing by transforming stored data into actionable decisions. In addition, when effectively implemented, it facilitates collaboration, improves information sharing, ensures security, and supports data recovery, capabilities that are increasingly vital in today’s business environment, where remote work and the use of personal devices are the new norm. A range of mature data storage methods exists, including in-memory storage, disks, cloud services, databases, file systems, and hybrid approaches. To enhance data access and system performance, data are typically stored in databases; for large-scale systems, decentralised databases are the preferred choice. Such storage not only preserves historical data patterns that inform current decision-making but also supports large-scale operations. For example, Google indexes trillions of web pages and performs more than eight billion searches per day by employing a persistent data storage architecture \cite{RN59, RN60, RN75}. Similarly, Meta (formerly Facebook) handles more than four billion likes per day, processing terabytes of new data with HDFS (Hadoop Distributed File System) on more than 100 petabytes of storage \cite{panda2022high}. Managing these vast and constantly updating datasets requires advanced storage architectures and high-performance computing resources. 

In general, the volume of data to be processed today is often too large for a single processor to handle efficiently. Consequently, there is a pressing need for parallel processing when data are stored at a single site or distributed processing when data are dispersed across multiple geographically diverse sites. In developing such resource-intensive applications, it is essential to consider both hardware and software infrastructures. Large-scale data storage architecture deployments differ markedly from traditional RDBMS (Relational Data Base management Systems). While conventional systems may require significant hardware investments, large-scale data solutions often depend more on advanced software resources, data management techniques, and redesigned network architectures to optimise performance. 

Other than storage, accessing distributed data is a serious concern, owing to the following reasons: simply storing the data is not enough, it is now mandatory to treat the distributed dataset as a whole, since accessing the data stored in just one of the available servers provides a partial view of the reality. In some cases, accessing them while they are in a desired format is also another essential consideration. Distributed data management systems have demonstrated promising results in resource-intensive applications. As pointed out in \cite{RN94, RN366, RN81}, distributed approaches can ensure efficient and secure data transfer and provide a framework for producing condensed versions of collected data. Evidently, a distributed approach to data management provides several key benefits, including:

\begin{itemize}
    \item \textit{Locality of data:} this is achieved by storing the data into servers which are close to the intended users. In turn, this means that the users can benefit from increased bandwidth and reduced latency, leading to efficient data transfers;
    \item \textit{High availability and fault tolerance:} if required, data may be replicated across multiple servers. This allow users to retrieve the data they are interested in from the server which is closest to them, and, in case of failure, no server is a single point of failure since the data can be retrieved by an alternative server, thus coping with network partitions and hardware failures seamlessly;
    \item \textit{Scalability:} adding more servers allows coping with the demand for handling more data, easily and transparently managing the increased load;
    \item \textit{Reduced costs:} a distributed database is typically more cost-effective than a centralised one for managing large amounts of data. This is because commodity hardware can be used in place of high-end, dedicated hardware. In practice, distributed databases rely on horizontal scaling (also called scaling out), i.e. the addition of new nodes/servers whilst centralised databases rely on both vertical scaling (also called scaling up) and sharding. Vertical scaling means upgrading the machines with better CPUs, faster memory, storage and network. Sharding refers to the practice of partitioning a huge centralised database into a number of smaller partitions, hosted on different servers (a form of horizontal scaling, done when needed to increase the performance);
    \item \textit{Flexibility:} The distributed nature allows more flexibility than a centralised solution, with regard to design choices.
\end{itemize}

Hence, nowadays the data management aspect is not being looked at as just a mere data dumping tool, but rather as a tool to support business analytics by first storing the data, and then extracting knowledge from it. Distributed data management is a key technology to enable efficient massive data processing and analysis. In the past, we had to manage only a limited amount of data, but, nowadays, we need to deal with a data deluge. The number of sources has increased over time, including retailers’ databases, logistics, financial and health data, social networks, pictures and movies taken from mobile devices, the IoT, scientific data, etc. 

This explosion is due to the mostly automated processes and the Internet-generated data, coupled with the ability to store almost all of the data (owing to the availability of cheap hard disks). The data contain both value and knowledge, so much that data are one of the key assets of modern companies: being able to extract useful information from data is critical for commercial exploitation. From a scientific perspective, scientists are at an unprecedented position where they can easily collect terabytes of information (and even more than that). However, in order to extract the knowledge, the data must be stored, managed, cleaned, and analysed. Hence large-scale data differ from typical scenarios as follows. First, the data are stored, but we have no time to read them all, so we read only a part of them (an approach which is sublinear in the time required). Second, the data are too big to fit in main memory, so we store them on disk (an approach sublinear in I/O) but throw some of them away (an approach sublinear in the space required). Third, the data are too big to fit in a single server, so in order to avoid throwing them away, we need to store and process them on multiple servers (a parallel or distributed setting, an approach sublinear in the communication required). This work aims to partly address the prevailing challenges; in this regard, the main contributions of this paper can be summarised as follows:

\begin{itemize}
    \item We propose an innovative approach to data storage, based on a distributed system using a P2P overlay and database management platform;
    \item We define requirements for developing a decentralised data storage tool in P2P network settings;
    \item We propose specifications of the IDSS that can manage distributed relational data based on the defined requirements;
    \item We propose methods for running complex distributed queries using DHT and approaches for integrating distributed data.
\end{itemize}

The rest of this paper is organised as follows. Section \ref{sec:background} provides background details
related to the technologies relevant for the current work. Section \ref{sec:proposed_arch} presents the proposed architecture, detailing the system description, methods to query the data, and considerations for implementation. We recall the state of the art related work in Section \ref{sec:related}. Lastly, we draw our conclusions in Section \ref{sec:conclusion}.

\section{Background}
\label{sec:background}

In this section we provide background details related to the technologies relevant for the current work. Table \ref{tab:abbreviations} lists the corresponding acronyms we use below.

\begin{table}[h]
    \renewcommand{\arraystretch}{1}
    \setlength{\extrarowheight}{1pt} 
    \caption{Acronyms used in this article.}
    \label{tab:abbreviations}
    \begin{tabular}{>{\arraybackslash}m{2.5cm}  p{0.25\textwidth}}
        \hline
        \textbf{\rule{0pt}{12pt}Acronym} & \textbf{Definition} \\
        \hline
        ACID & Atomic, Consistent, Isolation, and Durable \\
        API & Application Programming Interface \\
        AWS & Amazon Web Services \\
        BASE & Basically Available, Soft state, Eventual consistency \\
        CAP & Consistency, Availability, Partitioning \\
        CPU & Central Processing Unit \\
        DBMS & Data Base Management System \\
        DHT & Distributed Hash Table \\
        ETL & Extraction, Transformation and Load \\
        GCP & Google Cloud Platform \\
        GFS & Google File System \\
        HDFS & Hadoop Distributed File System \\
        IDSS & InnoCyPES Data Storage Service \\
        I/O & Input and Output \\
        InnoCyPES & Innovative Tools for Cyber-Physical Energy Systems \\
        IoT & Internet of Things \\
        KBR & Key-Based Routing \\
        ML & Machine learning \\
        MPP & Massively Parallel Processing \\
        NoSQL & Not Only SQL \\
        ODBC & Open Data Base Connectivity \\
        P2P & Peer-to-peer \\
        RDBMS & Relational Data Base Management System \\
        SDK & Software Development Kit \\
        SQL & Structured Query Language \\
        TTL & Time-to-live \\
        UQI & Uniform Query Identifier \\
        \hline
    \end{tabular}
\end{table}

\subsection{Data Storage Architectures}
\label{subsec:architectures}

There exist several techniques, methods, and technologies to store, retrieve, and manage data at fundamental levels. Classical approaches to data storage service implementation are either centralised or hierarchical. Their main drawbacks are scalability and single point of failure as the scale (in terms of the number of instances) rapidly increases. Indeed, the centralised approach represents a bottleneck not only in highly dynamic environments, in which the availability of resources and related features change frequently over time. The hierarchical approach, being based on a structured topology, provides better scalability and fault tolerance than the centralised approach, but the process of updating the information from leaf nodes to the root may be expensive. For these reasons, a decentralised, distributed approach to the data storage service is well suited. Below we give a summary of these approaches and theirs strength and shortfalls.

\subsubsection{Centralised and Decentralised Database Concepts}
\label{subsubsec:central_vs_decentral}
Storing data provides reliability for companies and therefore allows them to minimise downtime and protect precious data. The storage infrastructure can be developed based on a centralised, decentralised, or distributed system. A centralised system follows a client-server approach in which there is a central repository, the server, and other clients which feed off it. Even though this approach is simple and can minimise the cost to setup and maintain, it can also be problematic and create bottlenecks - especially if the traffic is high and many clients need to be served simultaneously \cite{RN309, RN405}. Besides that, if the server is down or suffers a denial-of-service attack, the service goes entirely off and becomes inaccessible. 

The decentralised approach improves the situation by preventing a single point of failure, allowing the replication of several repositories or servers \cite{RN278}. Decentralisation may be achieved by different means, such as a master-slave, a hierarchical, or a fully decentralised P2P approach. Both centralised and decentralised approaches often use vertical scaling and, with hardware constraints, there is a limit to the amount of scaling that can be achieved \cite{RN69}. The distributed system approach allows for vertical and horizontal scaling, thereby addressing fault tolerance, scaling, and system agility. Additionally, the node may need to apply a consensus algorithm to maintain its consistency. As a result, the system has low latency in response, and the failure of any load has little impact on the entire system \cite{RN69}. Their scalability and robustness are among the main reasons why it is recommended that systems adopt decentralised/distributed approaches to handle data storage.

In database systems, distributed databases provide seamless access to systems. They are made of multiple nodes where data is distributed across each computer that makes up the database cluster. Some common examples include \textit{Google Spanner}, \textit{Azure Cosmos} and all data warehouses. The motivation underlying the use of distributed databases varies between businesses; in general, when the collected data to be stored cannot be accommodated into a single computer system, then a distributed database approach could be applied. In some cases, when computation related to stored data takes longer than it should, a distributed approach can also be applied. Other motivating reasons include resilience, fault tolerance, and flexibility. In practice, distributed databases are the preferred choice over traditional databases, mostly because the traditional databases are challenged by the \textit{modern data} processing demands. Also, managing modern data demands an integration of several tools and methods \cite{RN94}. 

\subsubsection{Peer-to-peer Systems}
\label{subsec:p2p}

P2P exploded in popularity through file-sharing applications, and has since evolved into a very flexible framework, which is gaining interest within the scientific community \cite{RN94, RN95}. Although it was originally designed for pragmatic file-swapping applications, P2P techniques can be deployed for distributed resource sharing. Due to the ever-increasing traffic on the Internet, the need for a diversified wealth of applications is also growing. Unfortunately, traditional client-server approaches are too resource-aggressive to meet these challenges. In P2P setups, peers can be both clients and servers; their role may change dynamically as required. Communication among peers can be either structured or unstructured, giving flexible design choices. 

P2P systems may come in three architectures: pure, hybrid, and super peer. With a pure P2P system, a node can join the system and start exchanging data with any other nodes, embracing load balancing and fault tolerance at a high level \cite{RN55}. In a hybrid P2P system, the control information is exchanged through a central server, whilst data flow takes place in a pure P2P fashion \cite{RN83}. This architecture addresses the management problem present in pure P2P architecture. The central server monitors all communication between peers and ensures the coherence of information. However, this system is still challenged by the central management data transfer. If the central server goes down, the system loses coordination abilities, affecting data flow changes and information coherence. Of course, existing applications will not be affected by a central server failure, as the data flow between nodes continues regardless of whether the central server is functional or not. The super peer architecture aims to alleviate the challenges raised in pure P2P and hybrid P2P architectures. It embeds a centralised topology into a decentralised system. In this way, the architecture achieves performance, load balance, manageability, and resilience \cite{RN55}. 


P2P systems can be based either on structured or unstructured overlays, which are application-level networks (built on top of a network, which is usually the Internet; their topology is independent from the underlying network) that connect any number of nodes, each representing an instance of a participant peer. Usually, the deployed overlay structure has a huge influence in many ways, including security, robustness, and performance \cite{RN82}.

In unstructured overlays (i.e. overlays in which no topology exists), peers select neighbours through a predominantly random process. This approach is generally simpler to implement, but is characterised by inefficient routing mechanisms and techniques used to limit congestion in the event of messages flooding \cite{RN83}. In this architecture, some of the peers could not receive one or more messages of potential interest. The first generation of P2P file sharing applications was built on top of an unstructured overlay, and the systems relied on lookups via a central server, which stored the locations of all data items. Only after looking up the location of a data item via the server was the data transferred directly between peers. Other approaches, e.g., \textit{Gnutella} \cite{ripeanu2001peer}, use a flooding technique, i.e., look-up queries are sent to all peers participating in the system until the corresponding data item or peer is found. Other implementations of unstructured P2P are \textit{Freenet} \cite{hong2000distributed}, \textit{FastTrack} \cite{liang2006fasttrack}, and \textit{Bittorrent} \cite{pouwelse2005bittorrent}. Gnutella and other unstructured overlay approaches proved to be inefficient in scaling.

Structured overlays, also known as Distributed Hash Tables (DHT), are characterised by the existence of a specific topology that imposes explicit constraints on which nodes may be neighbours of a given peer \cite{RN84}. They are based on a geometric design that tries to consider, simultaneously, different constraints such as:

\begin{itemize}
    \item The overlay must be fully connected for resiliency to peer failure;
    \item Peers must have uniform degrees to avoid load imbalance;
    \item There must be an efficient key-based routing algorithm;
    \item There must be distributed algorithms for efficient and easy construction and maintenance of the overlay routing state and topology.
\end{itemize}

Structured overlays provide Key-Based Routing (KBR), where messages addressed to any key will be incrementally routed towards an overlay node responsible for that key. Note that KBR is also possible in unstructured overlays, but with a less strict guarantee on object lookup \cite{RN84}. To deliver a message based on a key to its root node, each node forwards the message using a locally maintained routing table of overlay links. Structured overlays are characterised by greater complexity compared to unstructured ones. However, they provide better routing mechanisms ensuring that, if the content sought is available within the network, then it will be retrieved and presented to the user \cite{RN456}.

P2P networks have not been fully explored in the context of database management. Several implementations of P2P systems have been focusing on file sharing, some focused on sharing computations (e.g., SETI@home), and others in communication (e.g., ICQ). Platforms like BitTorrent, Gnutella, Kazaa and other implementations are equally limited in terms of database functionality. First, their focus is mainly on file sharing without a concrete content-based querying/searching facility. Second, they are single system that focus on one task, limiting their scalability when an extension to other applications/functions is required. 

Nevertheless, a few implementations already exist. For instance, a work by Calvanese et al., \cite{RN457}, proposed an approach for data management using a P2P Data Integration System. This work explores the evolution from centralised to decentralised data management systems, highlighting the limitations of traditional mediator-based data integration systems and the advantages of P2P architectures. Despite having autonomous peers that manage their own data and are interconnected through P2P mappings, allowing for decentralised and dynamic data management, the approach was limited in the following ways. First, inconsistency: the experiments show the possibility of data inconsistencies to arise from local data violations, contradictions between local and incoming data, or mutually inconsistent data from different peers. Second, preferences and trust management: the user did not have freedom to choose the kind of data they want to access or the data they trust. Third, authorisation and privacy management: the peers did not have means to ensure controlled access to data, nor there is a way to allow peers to specify privacy policies and manage data confidentiality. Fourth, model management: allowing complete peer autonomy was problematic, hence handling the dynamic nature of P2P systems where peers can join, leave, or become temporarily unavailable was impossible. There were no means to handle partial and incomplete data; hence, the authors recommended future research to address these challenges. To the best of our knowledge, no work advanced P2P data management picking up from Calvanese et al \cite{RN457}. Recent advances for P2P data management are mostly focusing on blockchain data management which is primarily focusing on file and transaction records. However, even blockchain faces challenges with confidentiality, performance, scalability, and faulty tolerance making it an unfavourable choice for large-scale data management.

Evidently, currently research does not provide proper database functionality on P2P infrastructures \cite{RN95}. Some of the issues that specifically have to be addressed while implementing P2P data management include the following: 

\begin{itemize}
    \item Data location: peers must be able to reference and retrieve data stored on other peers;
    \item Query processing: the system should discover and identify peers with relevant data and execute queries efficiently;
    \item Data integration: when data sources use different schemas or formats, peers should access the data seamlessly, preferably in their native representation;
    \item Data consistency: if data are replicated or cached, maintaining consistency across replicas is essential.
\end{itemize}


In a typical scenario for P2P data management, a query must reach all possible data storage service instances efficiently. Hence, taking into account a constraint on the amount of time available for processing the query, the solution could involve the use of a structured overlay for broadcasting the queries, but closely resembling a flooding approach, which is typically used in unstructured overlays. Both structured and unstructured approaches have pros and cons. However, owing to their limitations, focusing on structured overlays along with the availability of P2P Software Development Kits (SDKs) and platforms to assess their suitability taking into account the current requirements for managing modern data is an important research area.

\begin{figure*}
    \centering
    \includegraphics[width=1\textwidth]{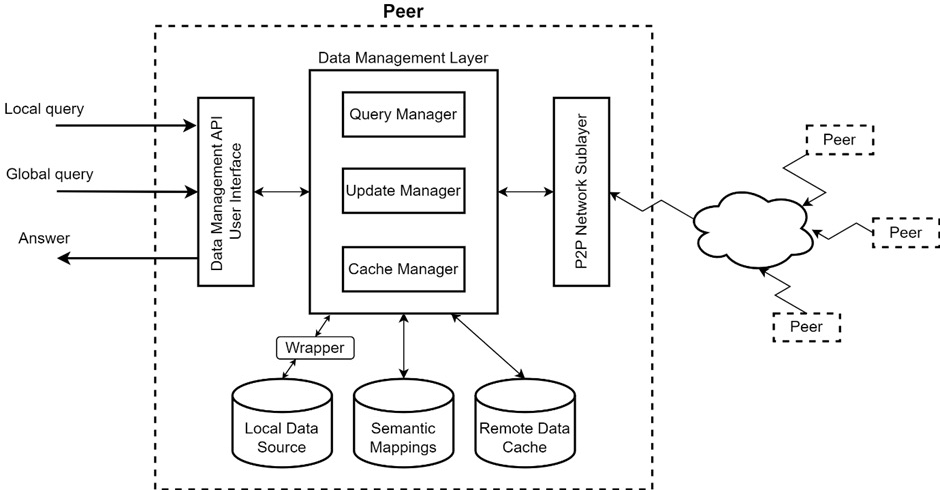}
    \caption{Peer reference architecture for data management \cite{RN95}}
    \label{fig:p2p_architecture}
\end{figure*}

Figure \ref{fig:p2p_architecture} provides a high-level description of a typical P2P data management platform. The design may be adopted in actual implementations by refining and arranging components as they fit specific requirements. The architecture presents three major components: (1) a querying interface, (2) a data management layer responsible for handling queries and primary database functions, and (3) a P2P overlay that handles all overlay functions. In cases where the P2P system implements a different database schema for each peer, a query manager (upon receiving a query request) retrieves semantic mapping information from the repository. The semantic mapping identifies peers in the system that contain data that are relevant to the submitted queries and reformulates the original query in terms that other peers can understand. Some systems can store semantic mapping functionality into a specialised peer. On the other hand, if all peers manage the same data schema, neither the semantic mapping repository nor the query reformulation functionality is required.

\subsection{Data Storage Tools}
\label{subsec:tools}

Data storage in computer systems is vital, however, it has been highly challenged by data isolation and difficulty in collaboration and sharing. In an effort to address these challenges, several database systems exist to offer ways to manage data in a consistent, stable, repetitive, and quick manner. 

Relational Data Base Management Systems (RDBMS) are among the popular data storage tools. Their architecture respects ACID properties to provide a means for data storage and allow more collaboration, reliability, security, and consistency \cite{RN62}. However, RDBMS fall short in managing modern data due to their limited scalability and rigidity in data schema. Data warehouse attempts to address some of the challenges experienced by RDBMS, specifically data analysis and reporting, and further allowing organisations to share their data \cite{RN74}. However, the warehouse approach soon faced the problem of a dead data zone (data that is no longer used or updated), limited collaboration, and difficulty in scaling \cite{RN157, RN74}. 

The evolution of big data, the technological advancement, and burst production of semi-structured and unstructured data further showed the limitation of relational data use and its difficulty in colossal data growth management. The reason is that scaling relational databases requires vertical scaling, and it suffers from hardware constraints, i.e. the number of physical devices that can be added. Partitioning can create a problem whilst joining tables and might lead to some discrepancies \cite{RN63}.

Non-relational databases, often referred to as Not only SQL (NoSQL) database also exist. They offer a schema-less handling of structured, unstructured and semi-structured data. Some NoSQL databases use Map-Reduce and HDFS, among others. They are often categorised between the ACID spectrum and Basically Available, Soft state, Eventual consistency (BASE), using the CAP (Consistency, Availability, and Partition) theorem and hence allowing horizontal scaling \cite{RN63, RN53}. Their architecture enables a database to seamlessly handle the scalability problem faced by relational databases using SQL. Furthermore, a data lake (a repository in which data are stored in raw format) allows all data to be stored and eliminates the data silo in the data warehouse \cite{RN54}. But a data lake can also lead to the problem of a data swamp (unmanaged data lake) with unfixed inconsistencies, duplication, and a poorly maintained lake. 

To handle data growth and challenges in lakes, many companies use cloud providers such as AWS, GCP, IBM, and Azure to handle their data. These providers allow for accessibility from anywhere and colossal scalability \cite{RN73}. However, the cloud can also lead to vendor lock-in, data dissipation, cost racking, and security challenges \cite{RN72}. Some businesses subscribe to the \textit{polynimbus} approach, which uses different clouds simultaneously, while others use a hybrid cloud system to address the challenge of using a single cloud provider \cite{RN71}. The ultimate choice of storage tool depends on the data to be stored, the flexibility, cost, security, and the organisation's demands. The \textit{polynimbus} approach, although helpful, may bring about complexities in data management.

As we are experiencing a rapid rise of data-supporting frameworks, technical experts have a wide range of choice to deploy their database infrastructure. These frameworks facilitate the creation and management of sophisticated cloud-based or local database systems (complex data storage systems). The frameworks can also provide the analysis of retrieved data \cite{RN90}. 

\subsection{Data integration strategies}
\label{sec:storage_and_integration}

The quest for new insights is driven by looking at other data sources that can solve the data discovery and new exploitation problems \cite{RN80}. Data integration involves strategies that allow one to get more out of data by combining them from different sources, in different formats, cutting across various fields and different time scales. This is usually not the case with single-sourced, locally stored data for a small or medium business. Usually, data are stored in a single server, generated and processed locally. In cases where the system must communicate with external services, APIs come to a rescue.

Given the sheer amount of data produced these days in heterogeneous forms and places, if data integration becomes flawed, it can lead to inconsistencies, inaccuracies, or even biased or wrong insights \cite{RN52, RN52}. To address these challenges, approaches based on the use of metadata can offer more information about the data, which can guide and allow us to understand data fusion \cite{RN351, RN119}. Similarly, federators such as Open data Base Connectivity (ODBC) enable users to query heterogeneous locations and get results from these sources \cite{feasel2020integrating}. ETL and ELT are other ways to combine data into a single standard format for querying \cite{RN119, RN155, RN157, RN119}. Other alternatives that use Resource Description Framework (RDF) for annotation can allow easy integration and cross-platform data extraction \cite{RN188}. However, these approaches come with their own challenges, such as increased computational overhead and potential performance bottlenecks when dealing with large-scale or real-time data integration. Additionally, ensuring data quality, security, and compliance across heterogeneous sources remains a significant hurdle, that requires a robust governance frameworks and efficient transformation techniques.

Leveraging \textit{Map-Reduce} (and its variants) and GFS are some of the ways in which large-scale distributed data have been brought together for analysis \cite{RN66, RN80}. However, these approaches rely on a centralised nature of resource coordination. Furthermore, because of the big data involved, there is still a need to look for innovative solutions that can handle the data fusion challenge in more effective ways. Indeed, limitations of data fusion approaches include the difficulty of dealing with data affected by error (lack of accuracy) and outliers. There is also the need to correctly handle conflicting data and correlated data. Another challenge is the intrinsic difficulty of managing high-dimensional data. In addition, many preprocessing steps are required, such as data normalisation and the imputation of missing data. Besides, a key challenge is related to correctly dealing with the semantics of data, especially for time-variant data.

\section{Proposed Architecture: IDSS Specifications}
\label{proposed}
We propose a distributed data storage tool leveraging a P2P overlay. Its design revolves around a query submission/management interface, a data manager and a P2P overlay. The architecture focuses on relational data, since this model is widely adopted and well understood. Therefore, the data storage service allows dealing with multiple relational sources in a fully decentralised way, and provides support for a user defined common schema (that can be customised by the user to accommodate different use-cases). Once defined, the schema must be the same on all peers.

\label{sec:proposed_arch}
\subsection{P2P overlay}
\label{subsec:overlay}
P2P networks support scalability, making them an appropriate choice for large-scale data management. To ensure that the requirements for data management are fulfilled, the design adopts a structured P2P overlay using DHTs. This is done to ensure that (i) the network is established and maintained, (ii) the peers are discoverable and new peers can join the network whilst allowing existing peers to leave the network, (iii) the data are distributed and can be searched using DHT lookup, and (iv) the communication among peers is smooth and consistent.

The adoption of the P2P architecture forces the way in which queries are handled. Whilst in a traditional centralised approach based on the client-server model, a synchronous approach can be used (the user submits a query and then synchronously waits for the server to execute the query and returns the results), the new P2P scenario requires an asynchronous approach in which the user submits a query to one of the available peers and, later, asks the peer for the corresponding results. 

Therefore, the IDSS relational schema includes a QUERY table, structured as follows (the actual implementation may of course differ, depending on the specific P2P implementation). 

\begin{itemize}
    \item \textit{id\_query:} it is the primary key field of the table. This field is significant locally for the peer node;
    \item \textit{uqi:} it is a Universal Query Identifier assigned to a submitted query; it has global scope across the P2P system: the same query on two different peer nodes is identified by the same value for this field on both nodes. This value is assigned when a query is submitted;
    \item \textit{value:} it is the SQL query string;
    \item \textit{arrival\_time:} it is the query submission time (a timestamp);
    \item \textit{TTL:} (Time To Live) it is the maximum time, in seconds, a node may use when processing a query;
    \item \textit{sender\_key:} it is the key of the node who forwarded the query during the broadcasting step. Clearly, this field is null for the initiator node;
    \item \textit{local\_exec:} allows distinguishing between queries that must be executed locally (value 0) and queries that have already been executed locally (value 1);
    \item \textit{completed:} allows distinguishing between queries for which merging of intermediate results must be performed (value 0) and queries for which the merging phase has been already performed (value 1);
    \item \textit{sent\_back:} allows distinguishing between queries for which delivering of results to the node specified in the sender\_key field has been performed (value 1) or not (value 0). Clearly, for a given query, this field is significant on all of the peer nodes that received the query, but not on the initiator node;
    \item \textit{failed:} its value is 1 if, for any reason and at any stage of processing, an abnormal condition has occurred which did not allow completing the execution of the query on the node. Note that, while a query can fail on a particular node, it may not necessarily fail on the other ones.
\end{itemize}

\subsection{Data and Query Management}
\label{subsec:data_query_mgt}
The proposed architecture adopts a relational database engine. This engine may be either a traditional database such as PostgreSQL, MySQL, Oracle etc., or an embedded database such as SQLite. Since an external database server represents yet another point of failure, for the sake of reliability and simplicity the IDSS will be based on an embedded database engine.

Regarding query support, IDSS must provide support for simple and complex queries with a) any condition on the values of the attributes, b) aggregate functions on its attributes (e.g., sum, average, min, max, etc.) and conditions, and c) nested queries.
A query life cycle includes the following steps: i) query submission; ii) broadcast; iii) execution and merge of the results.

\subsubsection{Query Submission}
\label{subsubsec:query_submission}
Traditionally, data is stored in data stores and managed by the DBMS. To retrieve data, users submit a query to the DBMS, which executes the query and sends the resulting recordset to the user. Under normal conditions, the user expects to receive the results relatively quickly, but, in a distributed environment, this is not the case. Query processing becomes more complex when databases are deployed in complex architectures and underlying networks such as a distributed P2P architecture. To efficiently execute a query in this architecture, the system must implement a query manager to facilitate effective and efficient execution. The query manager may follow a schema defined in section \ref{subsec:overlay} or design an alternative that fulfils other use-cases. 

To submit a query, the user contacts one of the nodes in an overlay using the IDSS client and submits the query of interest. The user can freely choose one of the available nodes; she needs to know only the contact information (hostname) for the data storage service instance. In this architecture, interactive synchronous execution of a query is not feasible. Indeed, the user cannot simply contact the service and wait (usually for a negligible fraction of time) for the results. Mainly because the query may not necessarily be performed only by the contacted node but could be performed potentially by almost all the nodes in the overlay. Since broadcasting the query and merging the results require a certain amount of time, the user specifies a \texttt{TTL}, which is the amount of time the user is willing to wait for the execution of the query (which is executed best-effort by IDSS, taking into account the constraint imposed by the \texttt{TTL} and the underlying network). Once a query has been submitted, the user will simply take note of the returned \texttt{UQI} to retrieve the results later.

\subsubsection{Query Broadcast}
\label{subsubsec:query_broadcast}
Upon receiving a query, a node broadcasts it within the overlay. Depending on the specific P2P overlay, the broadcast phase may give rise to duplicates, so that each node must rely on a mechanism for caching the queries to readily identify and discard a query already delivered to that node. In IDSS, this issue is dealt with by uniquely associating each query with a corresponding \texttt{UQI}. 

\begin{figure*}
    \centering
    \includegraphics[width=0.8\textwidth]{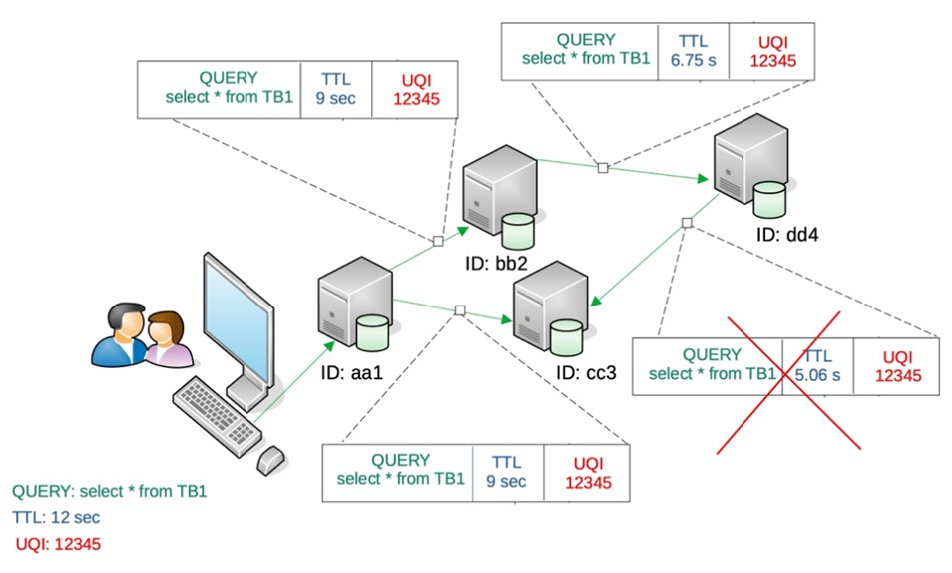}
    \caption{Query broadcast strategy}
    \label{fig:query_broadcast}
\end{figure*}

As shown in Figure \ref{fig:query_broadcast}, the receiver node checks its QUERY table to find an entry with the same \texttt{UQI} value; if the entry is found the query is discarded, otherwise the query \texttt{UQI} is inserted into the table and the node engages in the broadcast step by forwarding the query to its neighbour peers along with its \texttt{UQI} value. The query sent from node with ID \texttt{dd4} to node with ID \texttt{cc3} is discarded because it is a duplicate. In other words, the node with ID \texttt{cc3} receives the same query from nodes \texttt{aa1} and \texttt{dd4}, but the query from \texttt{aa1} arrived first and was already recorded, so the query from \texttt{dd4} is discarded.

\subsubsection{Execution and Merging}
\label{subsubsec:exec_and_merge}
\begin{figure*}[!t]
\centering
\subfloat[]{\includegraphics[width=2.5in]{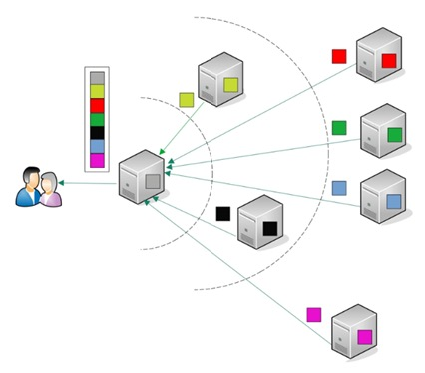}%
\label{fig:initiator_collector}}
\hfil
\subfloat[]{\includegraphics[width=2.69in]{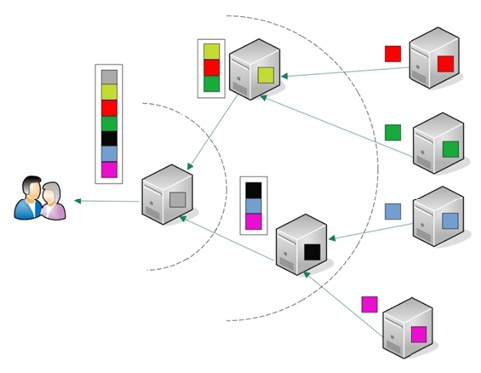}%
\label{fig:overlay_collector}}
\caption{Results collection and merging strategy. (a) Initiator node. (b) Intermediate node.}
\label{results_merging}
\end{figure*}

Regarding the query execution, each overlay node has in its cache table the entry related to the query submitted by the user, that must be executed locally as an intermediate step. It is worth recalling here that the overall result must be delivered to the initiator node, the one that received the initial query submission from the client. Regarding the intermediate local results stored on each involved node, two merging approaches are possible, as depicted respectively in Figure \ref{fig:initiator_collector} and Figure \ref{fig:overlay_collector}:

\begin{itemize}
    \item Each node performs the query locally and sends back its intermediate results to the initiator node, which, upon receiving them all (best effort), can then merge the results including its local data to produce the overall query result. This approach is simple and can be easily implemented; however, it puts all of the burden on the initiator node, resulting in a strong imbalance of the workload, besides the peak network traffic directed towards that node. See Figure \ref{fig:initiator_collector} for an illustration;
    \item Each node performs the query locally, receives the partial results obtained by all the nodes to which it directly forwarded the query, locally merges these results with its own (again, best effort), and finally sends back its intermediate result to the node from which it received the query. This approach, although more complicated than the previous one, ensures better load balancing within the network. An illustration is provided in Figure \ref{fig:overlay_collector}.
\end{itemize}

To implement the merge approach shown in Figure \ref{fig:overlay_collector}, we store in the cache of a node, besides the information related to a query, the identifier of the node from which the query was received in order to successfully perform the step of sending back the locally merged results. This information is of course meaningless on the initiator node, so in its table we store instead a sentinel value. 
 
To correctly merge the intermediate results, a node must wait a certain amount of time, to give a chance to the other nodes to which it forwarded the query to execute it and then send back the results. For this reason, when the user submits a query to the system, she can specify an initial \texttt{TTL} value. The \texttt{TTL} value is broadcasted along with the query and its \texttt{UQI} value, but its value differs at each hop and is computed as follows:

\begin{equation}
    \text{\(newTTL\)} = \text{\(oldTTL\)} \times \frac{3}{4}
    \label{eq:ttl_reduction}
\end{equation}

In practice, we decrement the \texttt{TTL} value at each hop by a constant factor, since at each hop the receiving node needs to wait for fewer nodes that will send back to it their intermediate results. The received results will be merged with its local data. Basically, we need to multiply the \(oldTTL\) value by a factor \(x\) such that \(0 < x < 1\) to obtain the \texttt{TTL} value to be used in a peer receiving a forwarded query. Setting \(x = \frac{3}{4}\) is a heuristic approach that we believe provides enough time for a peer to complete its merge operations and to forward the results back to the peer from which the query was received. Therefore, the whole merge process is best effort: a node waits from the query arrival time for at most the time specified by the \texttt{TTL}. 

Consequently, higher \texttt{TTL} values correspond to more accurate results returned for the query, since more information can be forwarded and processed. The user specifies the \texttt{TTL} value to be used for the submitted query, considering the trade-off involved: the lower the \texttt{TTL}, the faster the query response time and, correspondingly, the lower the overall quality of the query results (since some intermediate results provided by the nodes in the overlay will be silently discarded). Therefore, the \texttt{TTL} value should be directly proportional to the number of records expected to be retrieved (linear relationship): it should be increased for queries which are expected to retrieve a high volume of data and decreased for queries expected to return less data. It is worth recalling here that the total amount of data retrieved by a query depends both on the number of peers and on the number of records to be retrieved and stored on each peer.

\subsubsection{Query States}

To correctly manage the submitted queries, we model query execution by using the following states.

\begin{itemize}
    \item \texttt{QUEUED}: the query has just reached the node;
    \item \texttt{LOCALLY EXECUTED}: the query has been executed on the node and the local recordset (i.e., the set of records retrieved by the query, also known as resultset) is available;
    \item \texttt{COMPLETED:} the merge phase is done and a merged resultset is available;
    \item \texttt{SENT BACK:} merged results have been sent back to the node that originally forwarded the query during the broadcast phase;
    \item \texttt{FAILED:} an error occurred when executing the query. 
\end{itemize}

\begin{figure*}[h]
    \centering
    \includegraphics[width=0.8\textwidth]{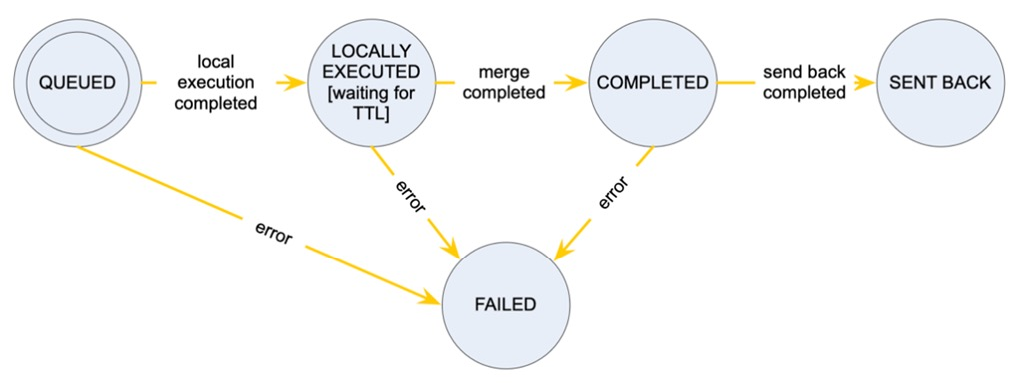}
    \caption{Query states on an overlay node.}
    \label{fig:querystate_overlay}
\end{figure*}

\begin{figure*}
    \centering
    \includegraphics[width=0.8\textwidth]{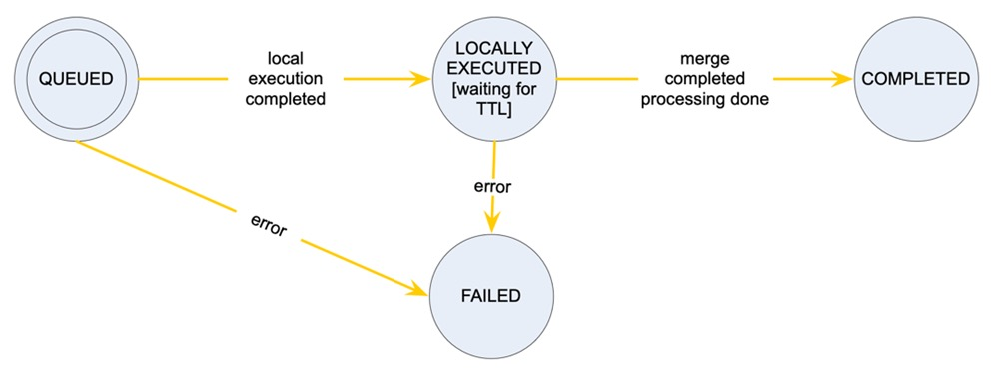}
    \caption{Query states on the initiator node.}
    \label{fig:querystate_initiator}
\end{figure*}

Figure \ref{fig:querystate_overlay} depicts the query states on an overlay node, whilst Figure \ref{fig:querystate_initiator} depicts the query states on the initiator node.

\subsubsection{Support for other types of queries}
Regarding the merging phase, we consider that the user can submit different kinds of queries to the system. For example, she can submit a simple \texttt{SELECT} query with any condition on the values of the fields (e.g., "\texttt{SELECT * from tb\_cpu\_dynamic where load > 0.1;}"). When performed locally, a query of this type returns a recordset with the records that meet the specified conditions. In the merge phase, a node appends the intermediate results it received to its local recordset, to obtain a single recordset matching the query. Therefore, the initiator node after merging the received recordsets with its own will have the overall query results.

However, the user can submit to the system queries with aggregate functions on its fields, such as sum, average, maximum, minimum and count, together with some conditions specified on the fields (e.g., "\texttt{SELECT sum (load) from tb\_cpu\_dynamic where load > 0.1;} "). When a query of this type is executed locally on a node, it generates a recordset with just a single record (for the aggregate function), containing the local results. Results are merged as outlined above, but the initiator node is then required to post-process these results by applying the corresponding aggregate function. 

This works without problems for the maximum, minimum, sum and count operators, since these operations provide the same result for a dataset handled by just a single node or for the same dataset partitioned among two or more nodes. It is worth noting here that, unfortunately, this is not true for the average operator, and we need to adopt appropriate strategies to deal with queries that contain the average operator. 

Given the definition of arithmetic mean, a simple way to circumvent this problem is to replace, at each node, the average function that appears in the query with the sum and count functions applied to the items before running the query locally (e.g., the query: "\texttt{select avg (load), max (load) from tb\_cpu\_dynamic;}"  becomes: "\texttt{SELECT sum (load), count (load), max (load) from tb\_cpu\_dynamic;}"). The initiator node applies the maximum, minimum, sum and count functions as discussed to compute the overall result. It deals with the average function by replacing each pair of introduced sum and count fields with the corresponding average value (sum divided by count).

\subsubsection{Nested Queries}
The IDSS design provides support for nested queries. A nested query is a \texttt{SELECT} query that is nested inside a \texttt{SELECT}, \texttt{UPDATE}, \texttt{INSERT}, or \texttt{DELETE} SQL query, e.g.: "\texttt{SELECT Model from Product where ManufacturerID in (SELECT  ManufacturerID from manufacturer where manufacturer = 'Dell');}".

With nested queries the level of complexity can vary greatly, considering that it is possible to have a deep hierarchy of nesting in a single query, a subquery predicate may contain a reference to a column of a table of a query ancestor, it is possible to include aggregate functions, etc. Therefore, handling nested queries in distributed systems is in general not a simple task. To cope with this additional complexity, by design IDSS limits the hierarchy to two levels of depth, allowing a parent query and an arbitrary number of nested subqueries, which in turn cannot contain additional nested subqueries. 

Each child can deal with a subquery single field or a single aggregate function; there can be no correlation between the parent query and the subquery. Moreover, the implementation will not deal with the case of both the parent query and subquery containing aggregate functions; finally, nested subqueries must also be of the same kind, i.e., all of them process a normal field, or contain an aggregate function. Despite these limitations, the system can handle a large range of queries that can be meaningful to the user, the simplest example of which would be: "\texttt{SELECT * from tb\_cpu\_dynamic where load > (select avg (load) from tb\_cpu\_dynamic);}" to obtain dynamic information on the CPU whose current load is greater than the average of the loads of all the available CPUs.

\begin{figure*}
    \centering
    \includegraphics[width=1\linewidth]{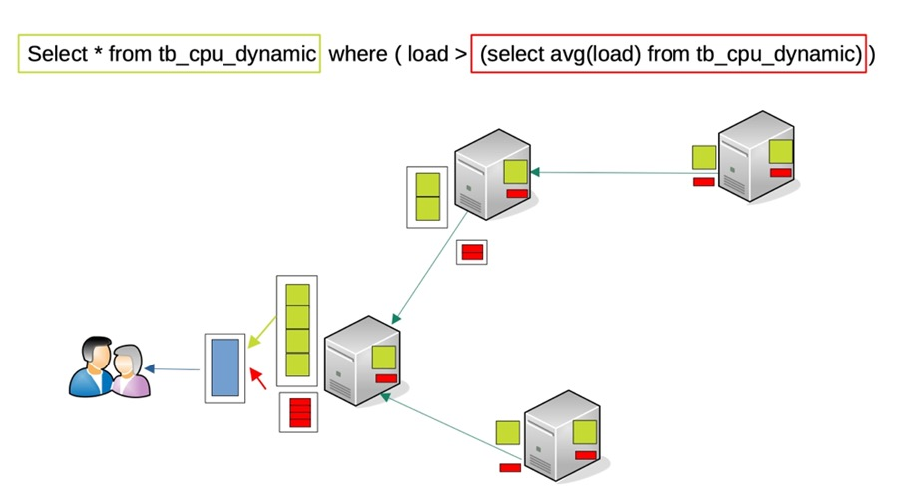}
    \caption{Execution of a nested query with an aggregation function.}
    \label{fig:nested_queries}
\end{figure*}

The result of simple queries containing aggregate functions is available only on the initiator node. It follows that the execution of queries with nested subqueries that use aggregate functions cannot be done independently on each node due to the aggregation functions used, and requires more processing with regard to simple queries. As shown in Figure \ref{fig:nested_queries}, specifically, each node must partition the query into elementary clauses, each to be run independently as in the case of simple queries: the parent query (without considering the where clause) and the various subqueries. The parent node of the query produces a recordset on the initiator node broader than that produced by the initial query, while the result of the aggregate function contained in each subquery will only be available on the same node. At this point, the initiator node owns all the information needed to retrieve the final recordset.

\subsection{Implementation}
\label{subsec:other_info}
The proposed design has two major technical components, the P2P overlay and a DBMS. Given the existing technologies, there exists a wide range of choices for the implementation. Since the proposed design demands that the data storage service must be bundled as a self-contained server application that does not rely on external services, it is mandatory for both the P2P infrastructure overlay and the DBMS to be used as libraries. Hence, the implementation of the database should rely on an embedded database engine. Other choices would follow specific requirements given a use-case, for instance, it is highly recommended to use the C/C++ programming language for the IDSS implementation, owing to the need to make IDSS as fast as possible, considering that IDSS must support multiple concurrent users. In particular, it is well known that C code runs significantly faster than most other compiled programming languages (with interpreted languages, such as Python, being slower than compiled ones and consequently unsuitable for enterprise-level server applications).

The P2P support is provided by D1HT \cite{D1HT-paper}, a novel distributed one-hop hash table which is able to maximise performance with reasonable maintenance traffic overhead even for huge and dynamic P2P systems. D1HT has reasonable maintenance bandwidth requirements even for very large systems, whilst presenting at least half the bandwidth overhead of previous single-hop DHTs. Support for relational data is provided by SQLite \cite{sqlite}, a small, fast, self-contained, high-reliability, full-featured, SQL database engine. The GRelC technology (Grid Relational Catalog) \cite{grelc} used in IDSS allows transparent and uniform access to different data sources. In addition, through its MultiQuery library, IDSS can easily handle a large amount of data in its relational database, with the ability to perform many  insert/update/delete queries one-shot.

IDSS is based on a Service Oriented Architecture (SOA), implemented through the use of Web Services (WS). In particular, IDSS leverages the gSOAP Toolkit \cite{gsoap} \cite{gsoap-conference} to provide both the server and its clients, based on SOAP (Simple Object Access Protocol) and XML (eXtensible Markup Language). IDSS makes extensive use of threads, through the pthreads library (POSIX Threads), not only for serving concurrently multiple users, but also internally to manage correctly and as quickly as possible its own modules and functionalities. The current version of IDSS is a traditional Unix daemon; in the future, we may also expose its services through a standard web server interface. The available command-line clients provide support for retrieving, inserting, modifying, and deleting data.

Security is an important requirement, encompassing several aspects (e.g. the interactions between a user and an IDSS server, peer communication during the establishment of an overlay and during data transfer). It is important to ensure that users and IDSS server peers are authenticated and that only legitimate peers participate in an overlay. Database (data) security requires methods to ensure that the confidentiality, integrity, and availability of database systems are achieved at the highest possible rate \cite{RN88, RN93}. Briefly, IDSS implementation strives to protect data from unauthorised access through the use of TLS/SSL (Transport Layer Security/Secure Sockets Layer). In particular, X509v3 digital certificates are used to authenticate both users and servers in the IDSS overlay. If required, organisation specific authorisation policies can be added to the IDSS server (by modifying a provided SSL callback). 

For testing purposes, IDSS has been deployed on a virtual machine (VM) running Ubuntu 24.04.2 hosted on a MacBook Pro equipped with a 2.3 Ghz 8 core Intel Core i9 processor and 64 GB of RAM. The VM has been configured to use 4 cores and 32 GB of RAM. Different overlay sizes have been tested, ranging from 8 to 1024 peers, with the aim to verify basic and advanced features, memory footprint and query execution under different loads. The results show that IDSS exhibits a small memory footprint and is reactive even under high loads (considering the stress test conditions).

We are aware that a real deployment would require installing IDSS servers on separate machines, perhaps geographically spread; however, this is virtually impossible for obvious reasons. Rather than simulating the IDSS operations through a P2P simulator, we opted for the implementation of the actual server. Our hope is that the IDSS prototype could be useful not only to organisations and small/medium enterprises, but also to the scientific research community. Therefore, the IDSS source code is open source, and freely available for download\footnote{https://github.com/cafaro/IDSS} to those interested in exploring and modifying the provided functionalities.

\section{Related work}
\label{sec:related}

Here, we review related work, most of which has been developed in the context of grid computing. In \cite{mastroianni:2005} the authors adopt the superpeer model to design a P2P-based grid information service. A large-scale grid can be viewed as interconnecting small-scale networks, named Physical Organizations (POs). In \cite{puppin:2005},  a grid Information Service based on the superpeer model and its integration within the Open Grid Services Architecture (OGSA) \cite{foster:2002} is proposed. The system offers fast propagation of information and provides high scalability and reliability. The authors implemented this service complying to the OGSA standard using the Globus Toolkit 3. The overlay is based on superpeer nodes. 

DHT based grid information systems supporting multi-attribute range queries include SWORD \cite{Oppenheimer} and MAAN \cite{Cai}. They both rely on the use of multiple DHTs, one per attribute, and queries are routed identifying the DHT corresponding to the query preferred attribute. MAAN is based on Chord. The Mercury grid information service \cite{Bharambe} also provides support for range queries over multi-attributes. However, it is not based on DHTs, even though it keeps a separate logical overlay for each attribute of interest.

XPeer \cite{xpeer} is a self-organising P2P database whose purpose is to allow sharing and querying XML data, in particular with regard to the management of resource descriptions in grid environments.

A distributed model of grid resource information service that extends the information service model of the Globus Toolkit 4, compliant with WSRF (Web Services Resource Framework) and based on the superpeer model to support scalability and robustness in large scale systems was presented in \cite{huo:2007}.

A new approach \cite{Hao} to P2P-based grid information service imposes a deterministic P2P shape on the overlay based on the hypercube topology, which allows for very efficient query broadcasting. The authors also propose a transposition algorithm whose aim is optimise the overlay network's topology according to the access statistics between peers.

In \cite{Zhu}, a service architecture based on structured P2P topology is presented. In the proposed architecture, each resource node and Index server is given an \emph{m}-bit identifier by hashing resource attribute values. Then, each node connects with successor Index server according to its identifier. In order to avoid the structured P2P overlay breaking formal relationship of a virtual organisation (VO), all resource information are managed by a local Information server instead of the Index server. The Index server just works as an index to provide the resource query service. The authors also discuss the importance of range queries in Grid environments and observe that  structured P2P systems cannot solve this problem well. Therefore, to support range queries, they propose a tree data structure and Path Caching Schemes to improve their proposed architecture. Finally, through theoretical analysis and simulation, the authors  show that the proposed Grid information services architecture achieves the goals of efficiency, robustness, load balancing, and scalability. 

In \cite{Basu}, the authors present the design of NodeWiz, a grid information service allowing multi-attribute range queries to be performed efficiently in a distributed manner. This goal is attained by aggregating the directory services of individual organisations in a peer-to-peer information service. The nodes joining the NodeWiz overlay are periodically updated by information providers external to the overlay. The authors describe an algorithm to split the attribute space based on \emph{k}-means clustering and another one, called top-\emph{k} algorithm required to  order the nodes in NodeWiz according to their workloads and to identify the most overloaded one.  In order to make a multi-attribute range query efficient, these algorithm must be run periodically and the overlay is updated accordingly.

In \cite{El-Ansary}, the authors present an efficient broadcast algorithm for structured DHT-based P2P networks with minimal cost. In a system of \emph{N} nodes, a broadcast message originating at an arbitrary node reaches all other nodes after exactly \emph{N }- 1 messages. The algorithm is based on the perception of a class of DHT systems as a form of distributed \emph{k}-ary search. Efficiency is obtained by constructing a spanning tree that is utilised for  broadcasting messages.

PIndex \cite{Sahota} is a grouped peer-to-peer network  built on top of Globus MDS4. This system introduces peer groups in order to dynamically split the large grid information search space into many small sections with the goal of improving scalability and resilience. PIndex efficiency in message routing comes from concurrent routing of messages coupled with dynamic partitioning of the search space into sections.

Antares, the system proposed in \cite{Forestiero}, is  a bio-inspired algorithm allowing the construction of a decentralised and self-organised P2P information system. Based on the properties of ant systems, Antares exploits a number of ant-inspired agents for controlled replication and relocation of descriptors, documents  containing metadata information about Grid resources. The agents navigate the Grid through P2P interconnections, and replicate and spatially sort the descriptors in order to accumulate those represented by identical or similar indexes into neighbour Grid hosts. The authors refer to the information system as a self-structured one, owing to self-organising characteristics of ant-inspired agents and the way descriptors are associated with hosts, which is not pre-determined but continuously adapts to dynamic variations of the Grid. The authors observe that the self-structured organisation is able to combine the benefits of both unstructured and structured P2P information systems. 

Indeed, being basically unstructured, Antares is easy to maintain in a dynamic Grid, in which joins and departs of hosts can be frequent events. On the other hand, the aggregation and spatial ordering of descriptors can improve the rapidity and effectiveness of discovery operations, which is a beneficial feature typical of structured systems. Performance analysis proves that ant operations allow the information system to be efficiently reorganised thus improving the efficacy of both simple and range queries. 

Another proposal \cite{Tao} integrates a scalable DHT and ontology-based Information Service (DIS) for a grid system based on the following observations: (i) the churn phenomena requires strong self-organisation capacity  to maintain the rigid DHT structure; (ii) arranging a moderate identifier space for a DHT ring is extremely complex. As a consequence of a  large identifier space some nodes will be overloaded, while a small identifier space "will bring forth the same problem as the millennium bug". Taking these facts into account, and considering that  grid services are often described using XML (so that no support for semantic queries is available), the authors designed a scalable DHT- and ontology-based Information Service (DIS) for a grid system, which organises resources into a DHT ring based on the VO mode. In their design, the overlay is kept stable by forbidding volatile VO nodes to join as new nodes: they can only appear as a sub-domain of a stable VO node. The proposed ontology supports semantic-based information queries targeting grid resources. The authors show how this leads to a faster information query with improved  query precision.

PORDaS \cite{pordas} is a P2P database that provides location-transparent storage of data for grid applications. Written in Java, it is based on the FreePastry DHT and the Derby DBMS. Since queries are by default local, i.e. are performed on the local database only, global queries potentially involving  all other peers, explicitly require specifying in the query the global tables to be searched.

A system prototype for P2P query answering has been proposed in \cite{deductive}. The approach is based on declarative semantics for P2P systems and allows implementing a deductive database in which only those data that do not alter the local database making it inconsistent can be imported for neighbour peers and  Preferred Weak Models represent the consistent scenarios in which peers can import maximal sets of facts without violating integrity constraints. 

PeerDB \cite{peerdb} has been designed to allow distributed data sharing in P2P networks without a shared schema. It uses mobile agents at each peer and self-organising with regard to the management of subsets of peers that contain most information: those peers are kept in close proximity for direct communication. The design has been evaluated up to 32 peers.

OrbitDB\footnote{https://github.com/orbitdb/orbitdb?tab=readme-ov-file} is a recent development in the context of P2P databases. It is implemented on top of IPFS\footnote{https://ipfs.tech} for data storage and libp2p\footnote{https://libp2p.io} to sync automatically a database among the available peers. As a consequence, it is, by design, eventually consistent (i.e. the data stored into the peers' database may not be coherent at all times). OrbitDB provides support for events, documents, key-value and key-value indexed data. Its decentralised architecture exploits OpLog \cite{sanjuan}, which is an immutable, cryptographically verifiable and operation-based conflict-free replicated data structure for distributed systems. Two versions are available, respectively JavaScript and Go based.

Peerbit\footnote{https://peerbit.org} is also built around IPFS and libp2p, and targets mainly data-intensive applications such as live-streaming and cloud-gaming. It provides support for end to end encryption, and automatic sharding of the stored documents. Peerbit is implemented in JavaScript and TypeScript.

GUN\footnote{https://gun.eco} is a data syncing protocol, small and fast. It is a graph database that can also store key-value data, documents and relational data. Encryption and authorisation mechanisms are available, and the data can be stored into three different types of space: public, user and frozen space. In the public space, essentially no control is enforced, allowing anyone to insert, update and delete the data. In the user space, only the user who own the data can insert, update and delete them when the data are private; but, the data can be made public. Finally, in the frozen space data can only be inserted but not updated or deleted, and nobody owns these data.

DefraDB\footnote{https://open.source.network/defra-db} is a graph database built on top of libp2p. It is a document database, with support for encryption and data replication. in active replication, a selected peer is constantly receiving updates from the local node, whilst in passive replication, updates are broadcast to the peers without explicit coordination and user's intervention, by using a publish-subscribe protocol provided by libp2p. Therefore, the system is eventually consistent among the subsets of peers selected for passive replication.

The main differences with regard to IDSS and the previously described systems are (i) its use of a one-hop DHT instead of unstructured overlays or standard DHTs; (ii) its primary target, which is not grid computing but small and medium enterprises on the one hand and research on the other; (iii) IDSS can handle range queries easily, without the need for additional complicated data structures and caching schemes, using the query mechanisms discussed in the previous section. In particular, no additional algorithms are required for the overlay in order to perform efficiently complex queries such as multi-attribute range queries; (iv) IDSS relies on standard SQL, whilst some of the previous system leverages different models and underlying query languages, e.g. the use of an ontology to support semantic queries. Finally, to the best of our knowledge, these systems do not provide support for complex distributed queries in P2P networks, such as nested queries.

\section{Conclusion}
\label{sec:conclusion}
Data is growing, making storing the data cumbersome and challenging. Spreading the data across different storage can also lead to inconsistencies and inaccuracies if poorly managed. Keeping data can either follow a centralised approach with a single point of failure or a distributed system with low latency, ensuring scalability and high fault-tolerance. This paper has presented the design of a novel database architecture aiming to maximise system throughput and availability, avoiding issues related to a centralised setting. This has been done by providing support for complex distributed queries, with simple and efficient data integration. Hence, to recap, the design functionalities and specifications can be summarised as follows. IDSS is based on a P2P architecture and leverages an embedded relational database engine. A query is broadcasted and executed within the P2P overlay network, with results merged along the paths towards the  peer to which the query was initially submitted. IDSS provides support for asynchronous, distributed query management and for simple and complex queries, including aggregate functions and nested queries. Future developments include testing libp2p to exploit its Kademlia based P2P overlay. Even though this library is already available, its C++ implementation is still in progress, and some functionalities have not been released yet.

\clearpage

\bibliography{references}

\end{document}